\documentclass[journal=jacsat]{achemso}
\usepackage[T1]{fontenc} % Use modern font encodings

\usepackage{amsmath,color}
\usepackage{amssymb}
\usepackage{amsfonts}
\usepackage{amsthm}
\usepackage{mathtools}

\usepackage{algorithm}
\usepackage{algpseudocode}
\usepackage{dsfont}

\usepackage{bbold}
\usepackage{chemformula}
\usepackage{siunitx}

\usepackage{pdfpages}

\DeclarePairedDelimiterX\braket[2]{\langle}{\rangle}{#1 \delimsize\vert #2}
\DeclarePairedDelimiterX\braket3[3]{\langle}{\rangle}{#1 \delimsize\vert #2 \delimsize\vert #3}

\usepackage{accents}

\usepackage{fancyvrb}

\newcommand{\vE}{\mathbf{E}}

\newcommand{\ve}{\mathbf{e}}

\newcommand{\vmu}{\boldsymbol{\mu}}
\newcommand{\hmu}{\hat{\boldsymbol{\mu}}}

\newcommand{\vR}{\mathbf{r}}

\newcommand{\vk}{\mathbf{k}}

\newcommand{\vxi}{\boldsymbol{\xi}}

\newcommand{\hH}{\hat{H}}

\newcommand{\tr}[1]{\text{Tr}\left(#1\right)}

\newcommand{\avg}[1]{\left\langle #1\right\rangle}

\newcommand{\red}[1]{{\color{black} #1}}

	\title{Semiclassical Real-Time Nuclear-Electronic Orbital Dynamics for Molecular Polaritons: Unified Theory of Electronic and Vibrational Strong Couplings}
	
	\author{Tao E. Li}%
	\email{tao.li@yale.edu}
	\affiliation{Department of Chemistry, Yale University, New Haven, Connecticut, 06520, USA}
	
	\author{Zhen Tao}%
	\affiliation{Department of Chemistry, Yale University, New Haven, Connecticut, 06520, USA}
	
	\author{Sharon Hammes-Schiffer}%
	\email{sharon.hammes-schiffer@yale.edu}
	\affiliation{Department of Chemistry, Yale University, New Haven, Connecticut, 06520, USA}

\begin{document}
	
    \begin{tocentry}
        \centering
		\includegraphics[width=0.65\linewidth]{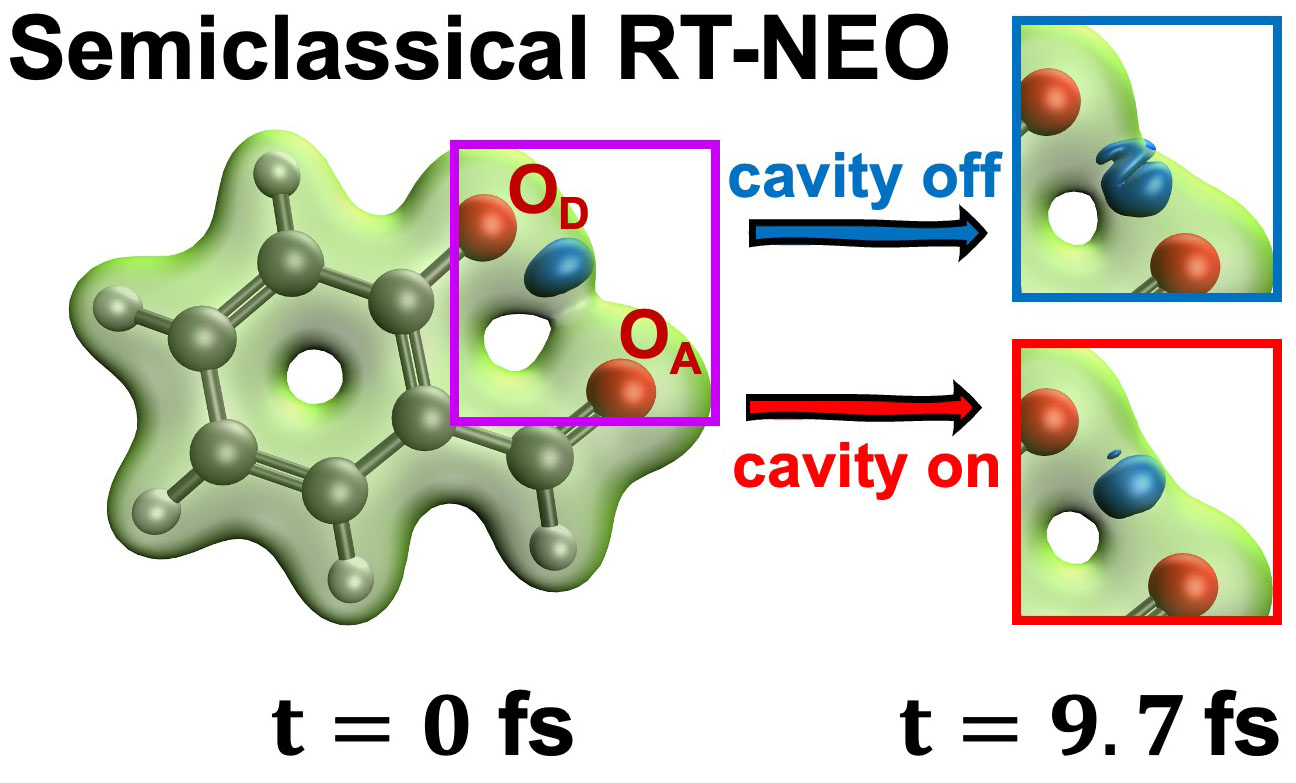}
	\end{tocentry}
	
	\begin{abstract}
		Molecular polaritons have become an emerging platform for remotely controlling molecular properties through  strong light-matter interactions. Herein, a semiclassical approach is developed for describing molecular polaritons by self-consistently propagating the real-time dynamics of classical cavity modes and a quantum molecular subsystem described by the  nuclear-electronic orbital (NEO) method, where electrons and specified nuclei are treated quantum mechanically on the same level. This semiclassical real-time NEO approach provides a unified description of electronic and vibrational strong couplings and describes the impact of the cavity on coupled nuclear-electronic dynamics while including  nuclear quantum effects. For a single \textit{o}-hydroxybenzaldehyde molecule under electronic strong coupling, this approach shows that  the cavity suppression of excited state intramolecular proton transfer is influenced not only by the polaritonic potential energy surface but also by the timescale of the chemical reaction. This work provides the foundation for exploring collective strong coupling in  nuclear-electronic quantum dynamical systems within optical cavities.
	\end{abstract}

	\maketitle
	
	\section{1. Introduction}
	
	Molecular polaritons, hybrid light-matter states stemming from strong light-matter interactions, \cite{Ribeiro2018,Flick2018,Herrera2019,Xiang2021JCP,Garcia-Vidal2021,Li2022Review} have attracted extensive experimental and theoretical attention due to the potential for modifying molecular properties. Examples of applications include controlling energy transfer, \cite{Coles2014,Zhong2017,Xiang2020Science} promoting electronic conductivity, \cite{Orgiu2015} and modifying photochemical \cite{Hutchison2012} and thermal \cite{Thomas2016,Thomas2019_science,Imperatore2021} chemical reaction rates. Molecular polaritons can form in different experimental setups, ranging from optical cavities, \cite{Deng2010} in which a large ensemble of molecules is  coupled to a confined photon mode, to plasmonic nanocavities, \cite{Pelton2019} in which a small number of molecules, possibly a single molecule, \cite{Santhosh2016} is coupled to a plasmonic mode. Depending on the frequency domain of the optical or plasmonic cavity mode, molecular polaritons can be mainly categorized as exciton-polaritons \cite{Deng2010,Keeling2020} or the recently discovered vibrational polaritons, \cite{Long2015,George2015} where a molecular electronic or vibrational transition, respectively, is strongly coupled to a cavity mode.
	
	In these electronic strong coupling (ESC) and vibrational strong coupling (VSC) domains, the intriguing experimental findings have been sparking intensive theoretical developments to enable the description of molecular polaritons. \cite{Ruggenthaler2014,Flick2017,Haugland2020,LiHuo2021,Luk2017,Li2020Water,Triana2020Shape,F.Ribeiro2018,Fregoni2018,Climent2019Plasmonic,Feist2020,Hoffmann2020,Tichauer2021,Yang2021QEDFT}
	Conventional quantum-optical theories of polaritons\cite{Jaynes1963,Tavis1968,Tavis1969} usually approximate the molecular electronic or vibrational transitions as two-level systems and the cavity mode as a harmonic oscillator. Thus, these approaches may fail to capture some important cavity effects of experimental interest, such as chemical bond formation and dissociation.  During the past decade, theories of molecular polaritons have been expanded to include the molecular details \cite{Ruggenthaler2014,Flick2017,Haugland2020,LiHuo2021,Luk2017,Li2020Water,Triana2020Shape,F.Ribeiro2018,Fregoni2018,Climent2019Plasmonic,Yang2021QEDFT} and the cavity mode structure beyond a single harmonic oscillator. \cite{Feist2020,Hoffmann2020,Tichauer2021} For example, under ESC, quantum-electrodynamical density functional theory (QEDFT) \cite{Ruggenthaler2014,Flick2017,Schafer2021}  extends time-dependent density functional theory (TDDFT), \cite{Marques2004} an efficient electronic structure method for calculating electronic excited states, to describe a  polaritonic potential energy surface. Under VSC, the recently developed classical cavity molecular dynamics (CavMD) \cite{Li2020Water,Li2021Collective} approach treats the infrared cavity mode as an additional "nuclear" coordinate, and a fully classical simulation of the coupled cavity-nuclear system has been shown to  qualitatively capture nonequilibrium dynamics under collective VSC arising from an ensemble of molecules coupled to the infrared cavity. \red{Other methods such as the exaction factorization \cite{Rosenzweig2022} and multiconfigurational time-dependent Hartree method (MCTDH) \cite{Triana2020Shape} also provide an accurate description of quantum effects under ESC or VSC.}
	In addition to a fully quantum or classical treatment of the coupled cavity-molecular system, a variety of semiclassical treatments of the light-matter system \cite{Sukharev2017,Chen2019Mollow,Chen2010,Yamada2018,Tancogne-Dejean2020,Bustamante2021,Schafer2021TDDFT} have also been shown to have the potential to accurately predict many light-involved processes from weak coupling to ESC. 

	Here, we report a novel semiclassical approach for describing molecular polaritons in which the quantum dynamics of the coupled nuclear-electronic system are self-consistently coupled to classical cavity modes. This approach is based on the recently developed real-time nuclear-electronic orbital TDDFT (RT-NEO-TDDFT) \cite{Zhao2020} approach, where both electronic and nuclear densities are propagated in real time. In contrast to full multicomponent TDDFT formalisms, \cite{Li1986TDDFTMulti,Marques2006TDDFT,Butriy2007}  typically only specified protons are treated quantum mechanically on the same level as all electrons. 
	Including both electron-electron exchange-correlation and electron-proton correlation effects, \cite{Chakraborty2008,Yang2017,Brorsen2017,Pavosevic2020} the linear-response \cite{Yang2018} and real-time \cite{Zhao2020} NEO-TDDFT approaches have been shown to produce reliable electronic and proton vibrational excited states. \cite{Culpitt2019JCTC, Culpitt2019JCP} These NEO approaches also include vibrational anharmonicity and nuclear quantum effects such as zero-point energies and proton delocalization. Moreover, the RT-NEO-TDDFT approach \cite{Zhao2020} has also been shown to directly capture the nonequilibrium coupled nuclear-electronic quantum dynamics of processes such as excited state proton transfer. 
	The semiclassical RT-NEO-TDDFT approach combines the classical motion of cavity modes with the RT-NEO-TDDFT dynamics of the molecular system and propagates the coupled dynamics self-consistently. We will show that this semiclassical approach not only captures the hallmark of ESC and VSC, namely the real-time Rabi oscillations and frequency-domain Rabi splittings, in a unified manner, but also provides a straightforward means for evaluating the cavity effect on coupled nuclear-electronic dynamics while including nuclear quantum effects.
	
	The main advantages of the semiclassical RT-NEO-TDDFT approach are the straightforward implementation, the capability to describe both the ESC and VSC domains simultaneously, and the potential scalability to collective strong coupling due to the use of classical cavity modes. Under ESC, although a semiclassical treatment of the coupled cavity-electronic system may not be as accurate as full quantum treatments such as QEDFT,
	\cite{Ruggenthaler2014,Flick2017} 
	\red{other work such as the semiclassical initial value representation (SC-IVR) method \cite{Miller2001,Cotton2013} has shown that treating electronic degrees of freedom as classical harmonic oscillators is valid for many scenarios. Due to the harmonic nature of cavity modes, which can be viewed as standing electromagnetic waves, a classical treatment is expected to be a good approximation. Moreover, even when the cavity modes are treated classically,}
	some quantum effects of the cavity modes can be recovered by introducing multiple trajectories in conjunction with the semiclassical algorithm.\cite{Li2020Quasi,Hoffmann2019Benchmark,Chen2019Mollow}
	\red{Additionally, in the limit of high excitations of cavity modes, a classical treatment becomes exact, whereas a quantum treatment usually requires greater computational cost to describe highly excited states.}
	Under VSC, compared with a fully classical treatment such as CavMD, \cite{Li2020Water} the inclusion of quantum protons is expected to be more reliable for probing ultrafast vibrational polariton spectroscopy. In the remainder of the paper, we introduce the fundamental theoretical concepts and central equations underlying our semiclassical approach and present illustrative examples. These applications include both a single \ch{HCN} molecule under ESC or VSC and the nonequilibrium dynamics of excited state intramolecular proton transfer.
	
	\section{2. Theory}
	
	\subsection{2.1. QED Hamiltonian under long-wave approximation}
	We start from a quantum-electrodynamical (QED) Hamiltonian for light-matter interactions: \cite{Cohen-Tannoudji1997,Flick2017,Li2020Origin}
	\begin{subequations}\label{eq:H_qed}
		\begin{align}
			\hat{H}_{\text{QED}} = \hat{H}_{\text{M}} + 
			\hat{H}_{\text{F}}.
		\end{align}
		Here, $\hat{H}_{\text{M}}$ denotes the conventional Hamiltonian for a molecular system composed of nuclei and electrons. This Hamiltonian is the sum of the kinetic and potential energies:
		\begin{align}
			\hat{H}_{\text{M}} = \sum_{i} \frac{\hat{\mathbf{p}}_i^2}{2 m_i} +  \hat{V}_{\text{Coul}}\left(\{\hat{\vR}_i\}\right) ,
		\end{align}
		where $m_i$, $\hat{\mathbf{p}}_i$, and $\hat{\mathbf{r}}_i$ denote the mass, momentum operator, and position operator, respectively, for the $i$-th particle (i.e., nucleus or electron), and $\hat{V}_{\text{Coul}}\left(\{\hat{\vR}_i\}\right)$ denotes the Coulombic interaction operator among all nuclei and electrons. 
		
		Under the long-wave approximation, the field-related Hamiltonian $\hat{H}_{\text{F}}$ is expressed as
		\begin{align}\label{eq:H_qed-3}
			\hat{H}_{\text{F}} = \sum_{k,\lambda}
			\frac{1}{2}\hat{p}_{k,\lambda}^2 + \frac{1}{2}\omega_{k,\lambda}^2\left(\hat{q}_{k, \lambda} + \frac{1}{\omega_{k,\lambda}\sqrt{\Omega\epsilon_0}} \hat{\vmu}_S\cdot \vxi_{\lambda}  \right)^2 .
		\end{align}
	\end{subequations}
	The cavity photon mode is characterized by the wave vector $k = |\vk|$ and the polarization direction $\vxi_{\lambda}$, which represents a unit vector satisfying $\vk \cdot \vxi_{\lambda} = 0$ (i.e., if the $\vk$ direction is $z$, then $\lambda$ is $x$ or $y$). This photon mode is linearly coupled to $\hat{\vmu}_S$, the total (electronic plus  nuclear) dipole moment of the molecular system. Here $\hat{p}_{k,\lambda}$, $\hat{q}_{k,\lambda}$, and $\omega_{k,\lambda}$ denote the momentum operator, position operator, and frequency, respectively, of the cavity photon. $\Omega$ denotes the effective volume of the cavity, and $\epsilon_0$ denotes the vacuum permittivity. %The field-related Hamiltonian in the form of Eq. \eqref{eq:H_qed-3} is also referred to as the Pauli--Fierz Hamiltonian.\cite{Flick2017}
	
	\subsection{2.2. Semiclassical approximation}
	
	We are interested in the semiclassical limit, where the cavity photons are treated classically. In this limit, the full QED Hamiltonian in Eq. \eqref{eq:H_qed} can be rewritten as
	\begin{subequations}\label{eq:H_semiclassical}
		\begin{align}
		\hH = \hat{H}_{\text{sc}} +  \sum_{k,\lambda}
		\frac{1}{2}p_{k,\lambda}^2 + \frac{1}{2}\omega_{k,\lambda}^2 q_{k, \lambda}^2,
	\end{align}
	where the semiclassical light-matter Hamiltonian is
	\begin{align}
		\hat{H}_{\text{sc}} = \hat{H}_{\text{M}}  +  \sum_{k,\lambda}\varepsilon_{k,\lambda} q_{k,\lambda} \hat{\mu}_{\lambda}.
	\end{align}
	Here, the light-matter coupling strength $\varepsilon_{k,\lambda}$ is defined as 
	\begin{align}
		\varepsilon_{k,\lambda} = \frac{\omega_{k,\lambda}}{\sqrt{\Omega\epsilon_0}}
	\end{align}
	and $\hat{\mu}_{\lambda} \equiv \hat{\vmu}_S\cdot \vxi_{\lambda}$ denotes the total (electronic plus nuclear) molecular dipole operator projected onto the direction of $\vxi_{\lambda}$.
	\end{subequations}

	Within this semiclassical treatment, the self-dipole term (i.e., the term proportional to $\hmu^2_S$) in Eq. \eqref{eq:H_qed-3} has been disregarded on the basis of the previous finding that neglecting this term is valid from weak  to strong coupling, \cite{Feist2020,Schafer2020} although it can fail under ultrastrong coupling. \cite{DiStefano2019,Schafer2020} There is no unique semiclassical treatment for the self-dipole term,\cite{Li2018Tradeoff} and identifying the optimal semiclassical form of this term in the ultrastrong coupling limit is beyond the scope of this manuscript.

	Given the semiclassical Hamiltonian defined in Eq. \eqref{eq:H_semiclassical}, we propagate the electrons and specified nuclei, typically protons, quantum mechanically and propagate the cavity photons classically. In particular, we propagate the dynamics of both electrons and quantum nuclei with the time-dependent Schr\"odinger equation:
		\begin{equation}\label{eq:TDSchrodinger}
		i\hbar \frac{\partial}{\partial t} \Psi(\mathbf{x}^\text{e}, \mathbf{x}^\text{n}; t) = \hH_{\text{sc}}(\mathbf{x}^\text{e}, \mathbf{x}^{\text{n}}; t) \Psi(\mathbf{x}^\text{e}, \mathbf{x}^{\text{n}}; t),
		\end{equation}
	where  $\mathbf{x}^\text{e}$ (or $\mathbf{x}^\text{n}$) denotes the collective spatial and spin coordinates of the electrons (or quantum protons), and the remaining heavy nuclei are assumed fixed.   This approximation is valid when the timescale of interest is smaller than the timescale of heavy nuclear motions, which is usually larger than tens of fs. On the other hand, we evolve the cavity photons according to the classical equations of motion:
	\begin{subequations}\label{eq:rt_photons}
	\begin{align}
		\dot{q}_{k,\lambda} &= p_{k,\lambda}, \\
		\dot{p}_{k,\lambda} &= -\omega_{k,\lambda}^2 q_{k,\lambda} - \varepsilon_{k,\lambda} \mu_{\lambda} - \gamma_{\text{c}} p_{k,\lambda}.
	\end{align}
	Here,  $\gamma_{\text{c}}$ denotes the cavity loss rate, which is introduced here to represent the imperfectness of the cavity mirrors. %, and $\mu_{\lambda}$ denotes the expectation value of the total  (electronic plus nuclear) molecular dipole moment projected along the direction of $\vxi_{\lambda}$.  
	In our simulations, we set the initial photon conditions as $q_{k,\lambda}(t=0) = p_{k,\lambda}(t=0) = 0$ and define $\mu_{\lambda}(t) = \avg{\hat{\mu}_{\lambda}(t)} - \avg{\hat{\mu}_{\lambda}(t=0)}$, where $\avg{\cdots}$ denotes the molecular expectation value, thereby setting  the molecular dipole moment to zero at $t=0$ and neglecting the effect of the permanent dipole moment on the cavity photons. This treatment is necessary to ensure that the cavity photons will not be excited at $t > 0$ without any external perturbation, i.e., a system starting in the ground state will always remain in the ground state in the absence of external perturbation; see Sec. Simulation Details for additional explanations.
	
	\end{subequations}

	\subsection{2.3. Semiclassical RT-NEO approach}
	
	Within the framework of the real-time NEO approach, \cite{Zhao2020} the nuclear-electronic wavefunction has the following form:
	\begin{equation}
		\Psi(\mathbf{x}^\text{e}, \mathbf{x}^\text{n}; t) = \Psi(\mathbf{x}^\text{e}; t) \Psi( \mathbf{x}^\text{n}; t),
	\end{equation}
	and the  time-dependent Schr\"odinger equation can be propagated separately for the electronic and nuclear components. Here, we  choose to propagate the von Neumann equations
	\begin{subequations}\label{eq:rt_NEO}
		\begin{align}
			i\hbar \frac{\partial}{\partial t} \mathbf{P}^{\text{e}}(t) & = \left[\mathbf{F}^{\text{e}}(t) + \sum_{k,\lambda} \varepsilon_{k,\lambda} q_{k,\lambda}(t) \hat{\mu}_\lambda^{\text{e}}, \ \mathbf{P}^{\text{e}}(t) \right] \\
			i\hbar \frac{\partial}{\partial t} \mathbf{P}^{\text{n}}(t) & = \left[ \mathbf{F}^{\text{n}}(t) + \sum_{k,\lambda} \varepsilon_{k,\lambda} q_{k,\lambda}(t) \hat{\mu}_\lambda^{\text{n}},\ \mathbf{P}^{\text{n}}(t) \right ]
		\end{align}
	\end{subequations}
	in the orthogonal atomic orbital basis.
	 The density matrices are defined as $\mathbf{P}^{\text{e}} = \mathbf{C}^{\text{e}}\mathbf{C}^{\text{e}\dagger}$ and  $\mathbf{P}^{\text{n}} = \mathbf{C}^{\text{n}}\mathbf{C}^{\text{n}\dagger}$, where $\mathbf{C}^{\text{e}}(t)$ (or $\mathbf{C}^{\text{n}}(t)$) denotes the coefficient vector of the electronic (or nuclear) wavefunction in the orthogonal atomic orbital basis.  $\hat{\mu}_\lambda^{\text{e}} = - |e| \sum_i \hat{r}_{i\lambda}^{\text{e}}$ and $\hat{\mu}_\lambda^{\text{n}} = |e| \sum_j Z_j \hat{r}_{j\lambda}^{\text{n}}$ denote the electronic and nuclear components of the molecular dipole moment, where $e$, $\hat{r}_{i\lambda}^{\text{e}}$, $Z_j$, and $\hat{r}_{j\lambda}^{\text{n}}$ denote the electronic charge, the $\lambda=x,y, \text{or\ } z$ component of the electronic position operator, the nuclear charge, and the $\lambda$ component of the nuclear position operator, respectively. $\mathbf{F}^{\text{e}}(t)$ (or $\mathbf{F}^{\text{n}}(t)$) denotes the \red{Kohn--Sham} matrix for the electrons (or nuclei) in the orthogonal atomic orbital basis\red{:
	 \begin{subequations}
	     \begin{align}
	         \mathbf{F}^{\text{e}}(t) &= \mathbf{H}_{\text{core}}^{\text{e}} + \mathbf{J}^{\text{ee}}(\mathbf{P}^{\text{e}}(t)) + \mathbf{V}_{\text{xc}}^{\text{e}}(\mathbf{P}^{\text{e}}(t))
	         -\mathbf{J}^{\text{en}}(\mathbf{P}^{\text{n}}(t))
	         -\mathbf{V}_{\text{c}}^{\text{en}}(\mathbf{P}^{\text{e}}(t), \mathbf{P}^{\text{n}}(t))
	         + \mathbf{V}_{\text{ext}}^{\text{e}}(t)\\
	         \mathbf{F}^{\text{n}}(t) &= \mathbf{H}_{\text{core}}^{\text{n}} + \mathbf{J}^{\text{nn}}(\mathbf{P}^{\text{n}}(t)) + \mathbf{V}_{\text{xc}}^{\text{n}}(\mathbf{P}^{\text{n}}(t))
	          -\mathbf{J}^{\text{ne}}(\mathbf{P}^{\text{e}}(t))
	         -\mathbf{V}_{\text{c}}^{\text{ne}}(\mathbf{P}^{\text{n}}(t), \mathbf{P}^{\text{e}}(t))
	         + \mathbf{V}_{\text{ext}}^{\text{n}}(t)
	     \end{align}
	 \end{subequations}
	 Here, $\mathbf{H}_{\text{core}}^{\text{e}(\text{n})}$ denotes the core Hamiltonian that includes the kinetic energy and the Coulomb interaction of the electrons or quantum nuclei with the classical nuclei;  $\mathbf{J}^{\text{ee} (\text{nn})}$ denotes the Coulomb interactions for the electrons or quantum nuclei;  $\mathbf{V}_{\text{xc}}^{\text{e} (\text{n})}$ denotes the exchange-correlation potential for the electrons or quantum nuclei; $\mathbf{J}^{\text{ne} (\text{en})}$ denotes the Coulomb interaction between the electrons and quantum nuclei; $\mathbf{V}_{\text{c}}^{\text{ne}(\text{en})}$ denotes the correlation potential between the electrons and quantum nuclei; $\mathbf{V}_{\text{ext}}^{\text{e}(\text{n})}(t)$ denotes the time-dependent external potential such as the light--matter coupling with the external pulse.}
	 Note that the \red{classical nuclei}  are treated as fixed classical point charges and do not contribute to the change in the molecular dipole moment relative to $t=0$.
	
	Eqs. \eqref{eq:rt_photons} and \eqref{eq:rt_NEO} form the working equations of the semiclassical RT-NEO approach of QED. We will use two molecular examples to illustrate the advantages and capabilities of this approach. For a single \ch{HCN} molecule with all electrons and the proton treated quantum mechanically, we show that our semiclassical RT-NEO-TDDFT calculation captures the real-time Rabi oscillations as well as the frequency-domain Rabi splitting under both ESC and VSC. For the  \textit{o}-hydroxybenzaldehyde (oHBA) molecule under ESC, with all electrons and the transferring proton treated quantum mechanically, our approach reveals the impact of the cavity on excited-state proton transfer dynamics. In the next section, we will provide additional simulation details.

	\section{3. Simulation Details}\label{sec:materials_methods}
	
	The semiclassical RT-NEO-TDDFT approach has been implemented in a developer version of Q-Chem. \cite{Epifanovsky2021} This approach entails propagation of Eqs. \eqref{eq:rt_photons} and \eqref{eq:rt_NEO}. We propagate the quantum molecular subsystem with a modified-midpoint unitary transform time-propagation scheme algorithm. \cite{Goings2018,Li2005} An additional predictor-corrector procedure \cite{DeSantis2020} is used to control the growth of numerical error during time propagation. The velocity Verlet algorithm is used to propagate the classical cavity mode. \red{The classical nuclei of the molecule are fixed at the specified geometry.} The outside cavity parameters are similar to those used in Ref. \cite{Zhao2020}. \red{The input files and plotting scripts are available at Github (\url{https://github.com/TaoELi/semiclassical-rt-neo}).}
	
	The initial conditions for propagating Eqs. \eqref{eq:rt_photons} and \eqref{eq:rt_NEO} are chosen to be the SCF ground state for the electronic and nuclear density matrices, $\mathbf{P}^{\text{e,n}}(t = 0)$, and $q_{k,\lambda}(t=0) = p_{k,\lambda}(t=0) = 0$ for the classical cavity mode. In Eq. \eqref{eq:rt_photons}, the evaluation of $\mu_{\lambda}$ requires additional explanation. As mentioned above, we define $\mu_{\lambda}(t) = \avg{\hat{\mu}_{\lambda}(t)} - \avg{\hat{\mu}_{\lambda}(t=0)}$, thereby neglecting the effect of the permanent dipole moment on the cavity photons  and ensuring that the cavity photons will not be excited at $t > 0$ without any external perturbation. 
	If we use an alternative strategy, in which $\mu_{\lambda}(t) = \avg{\hat{\mu}_{\lambda}(t)}$ (i.e., we consider the effect of the permanent dipole moment) and also set $q_{k,\lambda}(t=0)$ in a manner that satisfies $\dot{p}_{k,\lambda}(t=0)=0$ in Eq. \eqref{eq:rt_photons}, we can also ensure that the ground state will remain unchanged at $t>0$ in the absence of an external perturbation. This alternative strategy, which may contain an initial condition of $q_{k,\lambda}(t=0) \neq 0$ when the molecular permanent dipole is nonzero, can be understood physically in terms of a polarized photon field \cite{Mandal2020Polarized,Schafer2020} but is practically equivalent to our default treatment.

	For the DFT calculations, we use the B3LYP functional \cite{Lee1988,Becke1988,Becke1998} for electron-electron exchange-correlation and the epc17-2 functional \cite{Brorsen2017,Yang2017} for electron-proton correlation. When calculating the SCF ground state, a tight convergence criterion is needed to reduce the computational error in the real-time simulation (see SI for the input files). A time step of $\Delta t = 0.04$ a.u. is used for the real-time simulation.
	
	For the \ch{HCN} simulations, we use the cc-pVDZ electronic basis set \cite{Dunning1989} and an  even-tempered $8s8p8d$  protonic basis  set \cite{Yang2017} with  exponents  ranging  from $2\sqrt{2}$  to  32. Outside the cavity, at $t=0$, we apply a delta pulse to both the electronic and protonic Fock matrices, which can be expressed as $\mathbf{F}^{'\text{e,n}} + \mathbf{E}\cdot \vmu^{'\text{e,n}}$. Here, $\mathbf{E}  = (E_0, E_0, E_0)$ with $E_0 = 0.01$ a.u., and $\mathbf{F}^{'\text{e,n}}$ and $\vmu^{'\text{e,n}}$ denote the electronic or protonic Fock matrix and dipole moment vector matrix (in three dimensions) evaluated in the nonorthogonal atomic orbital basis, which are labeled with a prime superscript to distinguish from Eq. \eqref{eq:rt_NEO}, where the matrices are expressed in the orthogonal atomic orbital basis. Inside the cavity, we apply a delta pulse to the cavity mode (instead of the molecule) at $t = 0$: $q_c = q_c(t=0) + \Delta q_c$, where $\Delta q_c = 0.001$ a.u. under ESC and  $\Delta q_c = 0.3$ a.u. under VSC to increase the signal of the protonic dipole moment and reduce  the relative numerical error.
	
	We then calculate the power spectrum of the real-time dipole signal  by a Fourier transformation. Note that since we have only propagated a short range of time (50 fs), a direct Fourier transformation does not provide sufficient resolution in the frequency domain. In order to obtain enough resolution, following Refs. \cite{Bruner2016,Goings2018} , we take the Pad\'e approximation of the Fourier transform and calculate the power spectrum as follows:
	\begin{equation}\label{eq:FFT}
		P_e(\omega) = \sum_{i=x,y,z}  \lvert \mathcal{F}[\mu_i^e(t) e^{-\gamma t}] \rvert .
	\end{equation}
	Because the dipole signal calculated from  real-time time-dependent electronic structure theory does not contain any damping, a small damping term $e^{-\gamma t}$ is used (with $\gamma = 10^{-5}$ a.u.) to give an artificial linewidth of $1.7\times 10^{-3}$ eV = 13.8 cm$^{-1}$ to all of the peaks in the spectra.
	
	For the oHBA simulations, the cc-pVDZ electronic basis set is used in conjunction with a small $1s1p$ protonic basis set with an exponent of 4 to provide a qualitative description of the proton transfer reaction. The proton position is evaluated by $\tr{\mathbf{r}^{'\text{n}} \mathbf{P}^{'\text{n}}}$, where  $\mathbf{r}^{'\text{n}}$ and $ \mathbf{P}^{'\text{n}}$ denote the protonic position matrix and density matrix evaluated in the nonorthogonal atomic orbital basis. At $t = 0$, we model the \ch{S0$\rightarrow$S1} transition by enforcing a HOMO to LUMO transition in the electronic density matrix. In order to simulate proton transfer, we have added three additional proton basis function centers; see SI for the corresponding coordinates.
	
	\section{4. Results}

	\subsection{4.1. Electronic strong coupling}
	
\begin{figure*}
		\centering
		\includegraphics[width=1.0\linewidth]{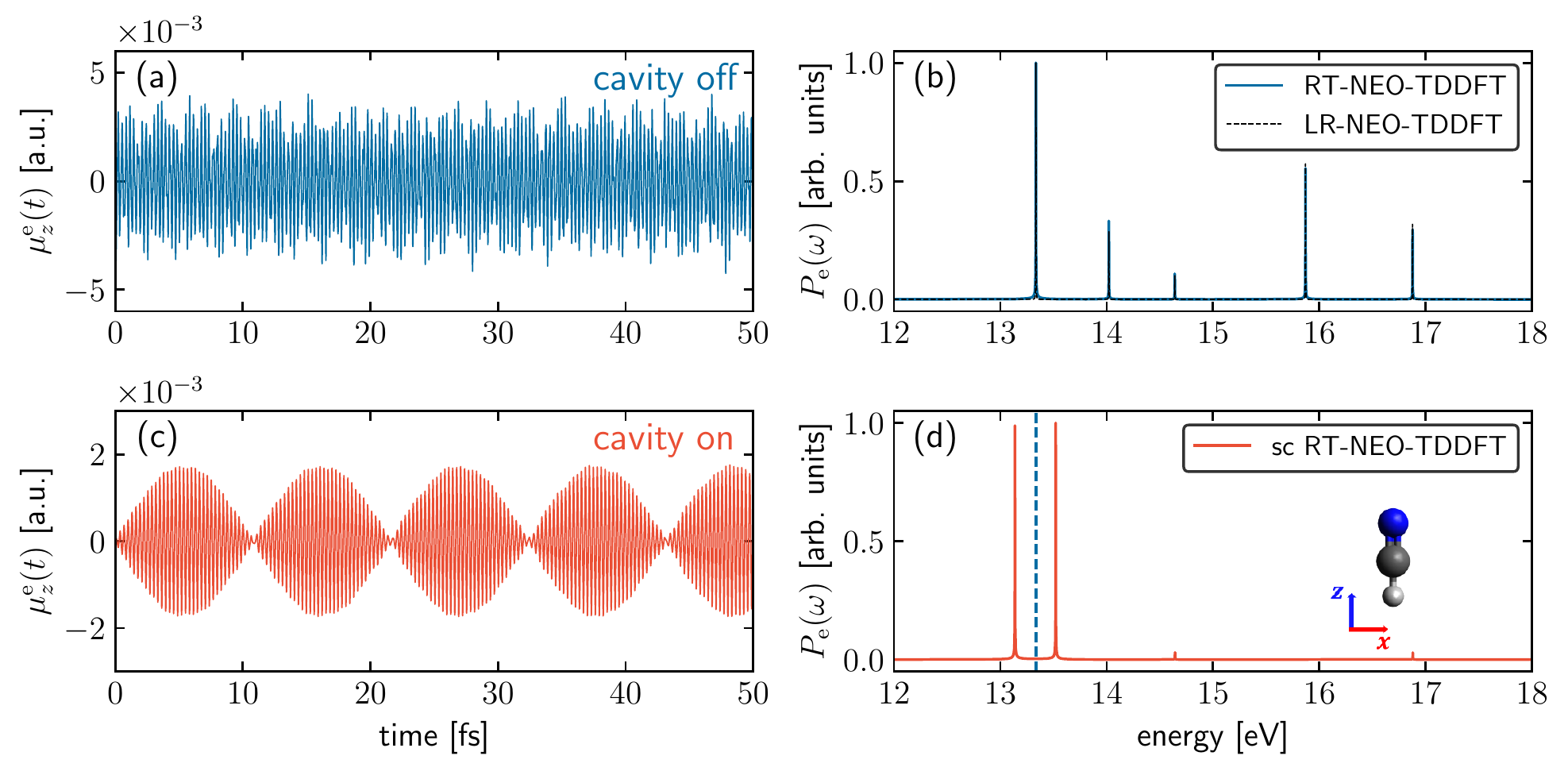}
		\caption{(a) Real-time NEO-TDDFT dynamics of the $z$-component of the \ch{HCN} electronic dipole moment, $\mu_z^{\text{e}}(t)$, in free space when the molecule is perturbed by a delta pulse. (b) Corresponding power spectrum, $P_\text{e}(\omega)$,  of the real-time dipole signal in (a) (blue line) as well as the linear-response NEO-TDDFT excited electronic state transitions (dashed black line). The real-time and linear-response spectra are virtually indistinguishable. (c) Semiclassical RT-NEO-TDDFT dynamics of $\mu_z^{\text{e}}(t)$ when the \ch{HCN} molecule is resonantly  coupled to a $z$-polarized cavity mode with frequency $\omega_\text{c} = 13.334$ eV, coupling strength $\varepsilon = 4\times 10^{-3}$ a.u., and no loss, and the cavity mode is perturbed by a delta pulse at $t = 0$. (d) Corresponding power spectrum, $P_\text{e}(\omega)$, of the real-time dipole signal in (c). \red{The geometry and orientation of the \ch{HCN} molecule are also depicted in (d).} The vertical dashed blue line denotes the cavity mode frequency, which is at resonance with an electronic transition in (b). Note that Rabi oscillations are observed in real time (c), and an UP and LP pair is formed in the frequency domain (d). All peaks in these  spectra have an artificial linewidth of $1.7\times 10^{-3}$ eV because a small damping term $e^{-\gamma t}$ with $\gamma = 10^{-5}$ a.u. is used to process the real-time signals.}
		\label{fig:ESC_dynamics}
	\end{figure*}

	Our first example is a single \ch{HCN} molecule oriented along the $z$ axis. Starting from the nuclear-electronic self-consistent field (SCF) ground state, this molecule is perturbed by a weak delta pulse at $t = 0$. Fig. \ref{fig:ESC_dynamics}a shows the real-time dynamics of the electronic dipole moment in the $z$ direction, $\mu_z^{\text{e}}(t)\equiv \avg{\hat{\mu}_z^{\text{e}}(t)} - \avg{\hat{\mu}_z^{\text{e}}(t=0)}$, outside the cavity for 50 fs. 
	%Note that for all the dipole moment dynamics shown in this manuscript, their values  have been rescaled to zero at $t = 0$ [$\mu_z^e(t) \equiv \avg{\hat{\mu}_z^e(t)} - \avg{\hat{\mu}_z^e(t=0)}$].
	Fig. \ref{fig:ESC_dynamics}b shows the corresponding electronic power spectrum $P_\text{e}(\omega)$ of the \ch{HCN} molecule (solid blue line). In order to check the validity of the real-time simulation, we also plot the electronic transitions calculated from linear-response NEO-TDDFT method (dashed black line). The relative heights of the linear-response peaks represent the corresponding oscillator strengths, where the maximum value is normalized to unity. The excellent agreement between the real-time and linear-response peaks (with a difference smaller than $10^{-3}$ eV) confirms the numerical stability of our real-time simulation.
	
	Next we simulate the real-time dynamics inside the cavity with a $z$-polarized lossless cavity mode at resonance with the strongest electronic peak in Fig. \ref{fig:ESC_dynamics}b ($\omega_\text{c} = 13.334$ eV) and the coupling strength set to $\varepsilon = 4\times 10^{-3}$ a.u. \red{The choice of a $z$-polarized cavity mode is to align with the largest transition dipole component among the three dimensions, where the geometry and orientation of the \ch{HCN} molecule are depicted in Fig. \ref{fig:ESC_dynamics}d.} Fig. \ref{fig:ESC_dynamics}c shows the dynamics of $\mu_z^{\text{e}}(t)$ when the cavity mode (not the molecule) is perturbed by a weak delta pulse at $t = 0$. In this case, real-time Rabi oscillations are observed due to the coherent energy exchange between the cavity mode and the molecular electronic transition. In the frequency domain, as shown in Fig. \ref{fig:ESC_dynamics}d, the original electronic peak at 13.334 eV (vertical dashed blue line) is split into two peaks, known as the lower polariton (LP) and the upper polariton (UP). These two polaritons are separated by a Rabi splitting of $\Omega_\text{R}=0.384$ eV.  The other electronic peaks in Fig. \ref{fig:ESC_dynamics}b are not significantly excited in Fig. \ref{fig:ESC_dynamics}d because the cavity mode frequency is far from these off-resonant electronic peaks, and therefore the excited cavity mode transfers only a small amount of energy to these electronic transitions. Figure S1 in the SI illustrates how the polaritonic spectrum depends on the cavity loss and light-matter coupling strength.

	\subsection{4.2. Vibrational strong coupling}
	
	\begin{figure*}
		\centering
		\includegraphics[width=1.0\linewidth]{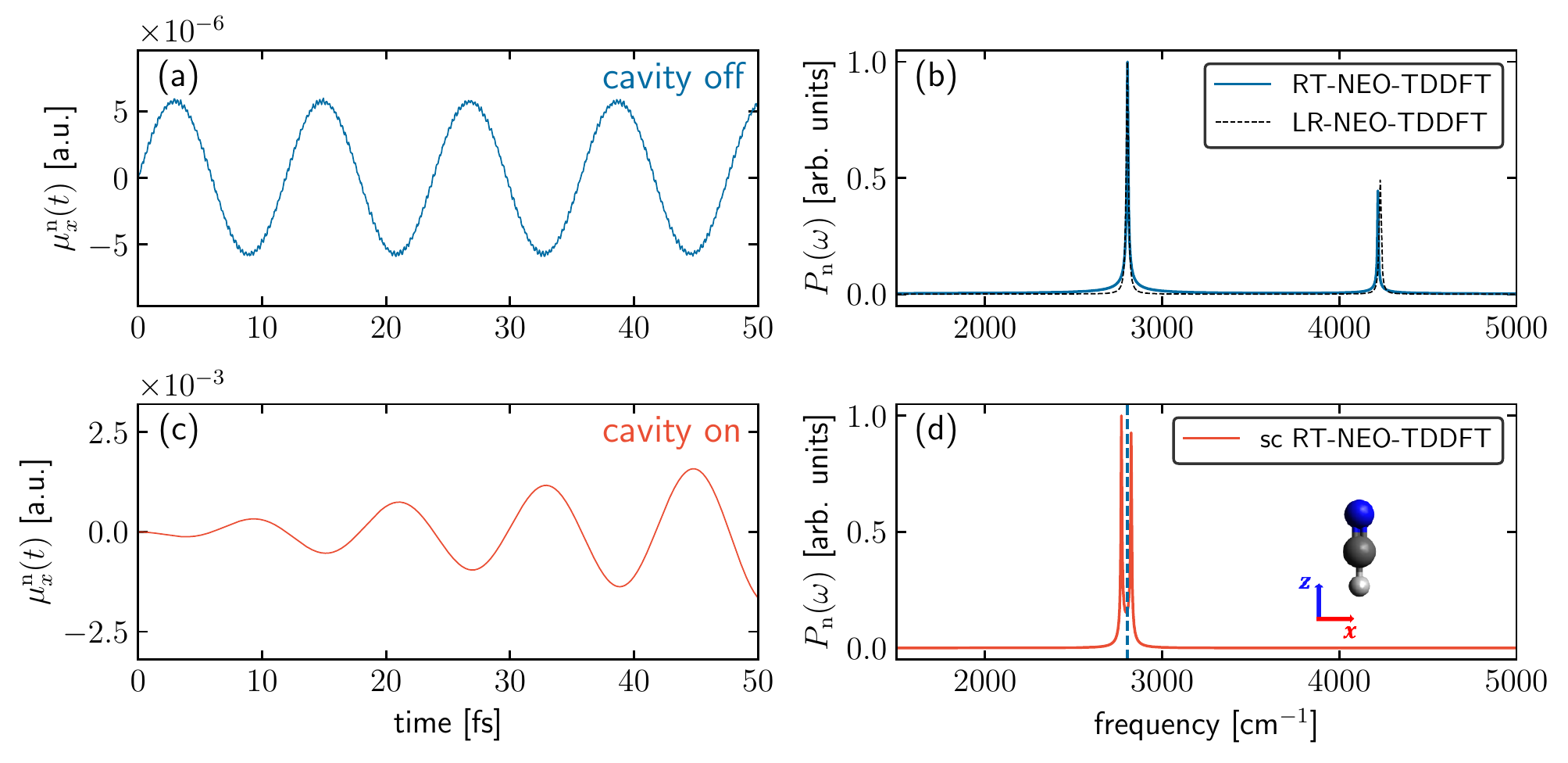}
		\caption{(a) Real-time NEO-TDDFT dynamics of the $x$-component of the \ch{HCN} nuclear dipole moment, $\mu_x^{\text{n}}(t)$, in free space when the molecule is perturbed by a delta pulse. (b) Corresponding power spectrum, $P_\text{n}(\omega)$, of the dipole signal in (a)  (blue line) as well as the linear-response NEO-TDDFT excited vibrational state transitions (dashed black line). The real-time and linear-response spectra are virtually indistinguishable. (c) Semiclassical RT-NEO-TDDFT dynamics of $\mu_x^{\text{n}}(t)$ when the \ch{HCN} molecule is resonantly coupled to an $x$-polarized cavity mode with frequency $\omega_\text{c} = 2803$ cm$^{-1}$, coupling strength $\varepsilon = 4\times 10^{-4}$ a.u., and no loss, and the cavity mode  is perturbed by a delta pulse at $t = 0$. (d) Corresponding power spectrum, $P_\text{n}(\omega)$, of the real-time dipole signal in (c). The vertical dashed blue line denotes the cavity mode frequency, which is at resonance with the lowest vibrational transition in (b). All peaks in these  spectra have an artificial linewidth of $1.7\times 10^{-3}$  eV (13.8 cm$^{-1}$) because a small damping term $e^{-\gamma t}$ with $\gamma = 10^{-5}$ a.u. is used to process the real-time signals.}
		\label{fig:VSC_dynamics}
	\end{figure*}

	In addition to the ESC domain, one important advantage of the semiclassical RT-NEO approach is the capability of treating the VSC domain. Again, we use the \ch{HCN} molecule oriented along the $x$ axis  as an example.  Outside the cavity, Figs. \ref{fig:VSC_dynamics}a,b show  the dynamics of the $x$-component of the nuclear dipole moment, $\mu_x^{\text{n}}(t)$ in the time domain and the corresponding power spectrum in the frequency domain. In Fig. \ref{fig:VSC_dynamics}b, the linear-response NEO-TDDFT peaks (dashed black line) are shown to be in very good agreement with the real-time NEO-TDDFT peaks. Specifically, the peak difference is 1 cm$^{-1}$ for the \ch{HCN} bending modes (left) and 12 cm$^{-1}$ for the \ch{HCN} stretching mode (right) with nearly identical relative oscillator strengths between the two peaks.
	
	Next an $x$-polarized lossless cavity mode is resonantly coupled to the \ch{HCN}  bending  mode  with cavity frequency $\omega_\text{c} = 2803$ cm$^{-1}$ and coupling strength $\varepsilon = 4\times 10^{-4}$ a.u. \red{For the \ch{HCN} geometry oriented along the $z$-direction, which is shown in Fig. \ref{fig:VSC_dynamics}d,  the doubly degenerate bending mode oscillates in the $xy$ plane.} Fig. \ref{fig:VSC_dynamics}c shows the dynamics of $\mu_x^{\text{n}}(t)$ for 50 fs after a weak perturbation of the cavity mode (not the molecule) at $t = 0$.  As the proton evolves much more slowly than the electrons, within 50 fs the molecule has not finished a period of Rabi oscillation, in contrast to the ESC case. In the frequency domain, however, as shown in Fig. \ref{fig:VSC_dynamics}d, the observation of a pair of polaritons with a Rabi splitting of $\Omega_\text{R}=55$ cm$^{-1}$ implies VSC.

	\begin{figure}
		\centering
		\includegraphics[width=1.0\linewidth]{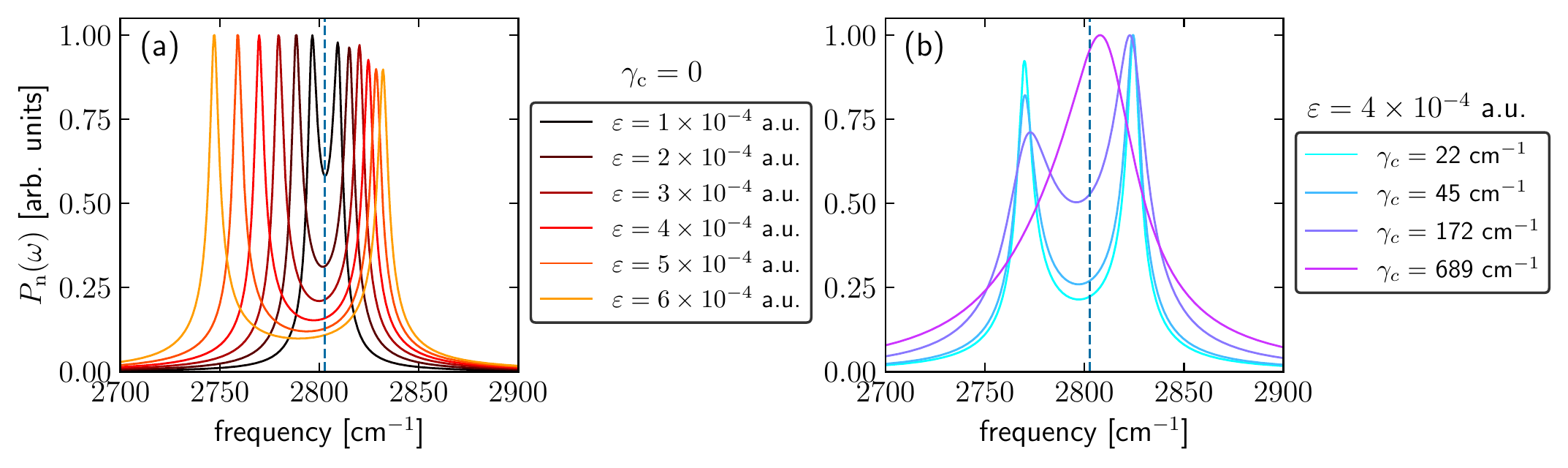}
		\caption{Power spectrum, $P_\text{n}(\omega)$, of the \ch{HCN} nuclear dipole moment when the \ch{HCN} bending modes are resonantly coupled to the $x$-polarized cavity with frequency $\omega_\text{c} = 2803$ cm$^{-1}$ (vertical dashed blue lines). (a) No cavity loss and the light-matter coupling strength ranges from $\varepsilon = 1\times 10^{-4}$ a.u. (black line) to $6\times 10^{-4}$ a.u. (orange line). (b)  The cavity loss ranges from $\gamma_{\text{c}} = 22$ cm$^{-1}$ (cyan line) to 689 cm$^{-1}$ (purple line) with the light-matter coupling strength set to $\varepsilon = 4\times 10^{-4}$ a.u. Increasing the cavity loss transforms the system from strong coupling (with a peak splitting) to weak coupling (with no peak splitting). All other simulation details are the same as for Fig. \ref{fig:VSC_dynamics}d. The molecular peaks have an artificial linewidth of 13.8 cm$^{-1}$ because of the small damping term used during signal processing. In part (b), the linewidths only partially arise from the signal processing and mostly originate from the coupling to the lossy cavity.}
		\label{fig:VSC_dependence}
	\end{figure}

	In Fig. \ref{fig:VSC_dependence}, we analyze the dependence of the polaritonic spectrum on the coupling strength and cavity loss \red{for the VSC case}. With all other parameters the same as in Fig. \ref{fig:VSC_dynamics}d, Fig. \ref{fig:VSC_dependence}a shows the polariton spectrum when the coupling strength between the lossless cavity and the molecule is tuned from $\varepsilon = 1\times 10^{-4}$ a.u. (black line) to $6\times 10^{-4}$ a.u. (orange line). As the coupling strength is increased, the Rabi splitting increases asymmetrically. Such asymmetric behavior, where the LP shifts more than the UP, is \red{not observed for HCN under ESC (see SI Fig. S1), most likely because the excitation is significantly higher. However, this asymmetry is also observed under ESC for the proton transfer system discussed below, which is also a relatively low-frequency excitation, as well as in a previous QED electronic structure calculation.\cite{Pavosevic2021} The origin of our asymmetry, in which the LP shifts more than the UP,  is currently unclear. In our treatment, the self-dipole term in the QED Hamiltonian has been neglected. Although Ref. \cite{Pavosevic2021} concludes that similar asymmetry arises from the inclusion of the self-dipole term in the QED Hamiltonian, Ref. \cite{Li2020Water}  suggests that including the self-dipole term would cause an opposite asymmetry, where the UP shifts more than the LP. The latter asymmetry agrees with the standard Hopfield model, \cite{FriskKockum2019} in which the $A^2$ term is included in the QED Hamiltonian. Thus, this issue warrants further investigation.}
	
	When the coupling strength is fixed as $\varepsilon = 4\times 10^{-4}$ a.u., Fig. \ref{fig:VSC_dependence}b shows the polariton spectrum when the cavity loss rate increases from $\gamma_{\text{c}}=22$ cm$^{-1}$ (cyan line) to 689 cm$^{-1}$ (purple line). When the cavity loss increases, the polariton linewidths increase until the Rabi splitting ($\Omega_\text{R} = 55$ cm$^{-1}$) disappears, illustrating a transition from strong coupling to weak coupling when the cavity loss increases. In Fig. \ref{fig:VSC_dependence}a, the linewidths arise from the signal processing, whereas in Fig. \ref{fig:VSC_dependence}b, the linewidths arise partially from the signal processing but  mainly from the coupling to the lossy cavity.
	
	\subsection{4.3. Proton transfer dynamics under electronic strong coupling}
	
	\begin{figure*}
		\centering
		\includegraphics[width=1.0\linewidth]{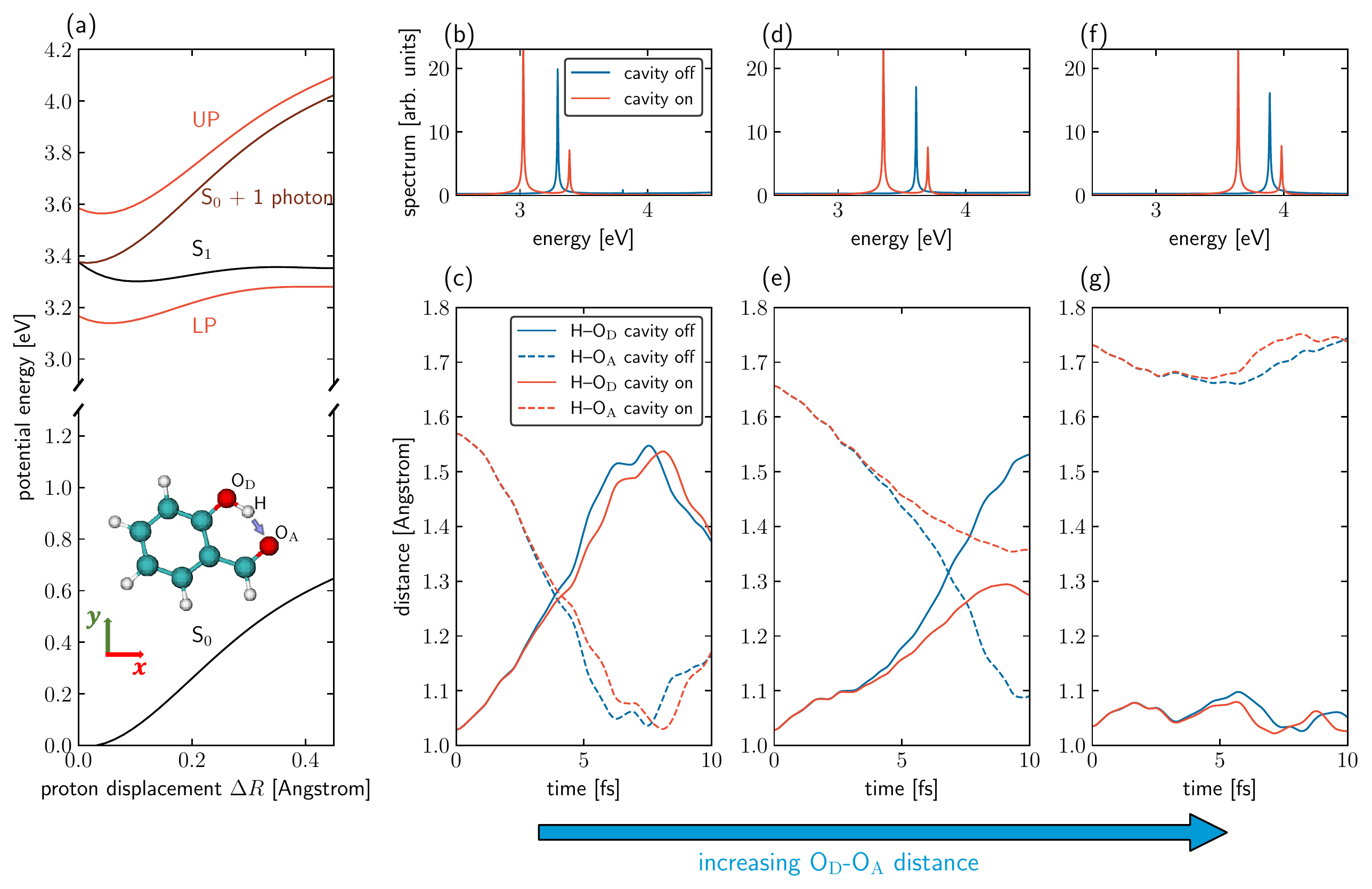}
		\caption{Excited state intramolecular proton transfer dynamics for an oHBA molecule under ESC. (a) Polaritonic potential energy surfaces as a function of proton displacement ($\Delta R$) for the oHBA molecule shown in the inset. The \ch{S0} and \ch{S1} states (black lines) are calculated by conventional LR-TDDFT. The minimum on \ch{S1} corresponding to the proton bonded to \ch{O_A} is $\Delta R = 0.45$~\AA. A pair of polaritonic states form (UP and LP, red lines) when the \ch{S1} state is near resonance with the \ch{S0} state dressed with a singly excited cavity photon (brown line); see Eq. \eqref{eq:H_polariton} for the effective Hamiltonian. (b) The corresponding power spectrum of the electronic dipole moment outside the cavity (blue line) or inside a lossless cavity (red line) when a $y$-polarized cavity mode is at resonance with the \ch{S0$\rightarrow$S1} transition (blue peak at 3.295 eV) with $\varepsilon = 4\times 10^{-3}$ a.u. (c) The corresponding excited-state intramolecular proton transfer dynamics outside (blue lines) or inside (red lines) the cavity. Solid (or dashed) lines denote the distance between the expectation value of the proton position and the donor (or acceptor) oxygen atom. (d,e) and (f,g) Two sets of spectra and proton transfer dynamics outside or inside the cavity when the distance between the oxygen atoms is increased. In Figs. (c), (e), and (g), the distance between the \ch{O_D} and \ch{O_A} atoms is 2.51 \AA, 2.57 \AA, and 2.64 \AA, respectively. When the proton transfer timescale (indicated by the crossing point of the red and blue lines) is smaller than the Rabi oscillation timescale ($\sim$ 11 fs), negligible cavity suppression of proton transfer is observed, as shown in (c), whereas increasing the proton transfer timescale while retaining a similar Rabi splitting leads to much greater cavity suppression of proton transfer, as shown in (d) and (e).}
		\label{fig:PT}
	\end{figure*}

	Beyond simulating real-time Rabi oscillations and the frequency-domain Rabi splitting, our semiclassical RT-NEO-TDDFT approach also provides a straightforward means  to probe nonequilibrium coupled nuclear-electronic dynamics under ESC or VSC conditions in a cavity. As an example, we consider the ESC effect on excited state intramolecular proton transfer (ESIPT) \cite{Scheiner2000,Aquino2005} for a single oHBA molecule inside the cavity. Unless otherwise specified, all of these calculations are performed for the restricted excited state geometry, which was obtained by optimizing the geometry in the electronically excited state with the distance between the proton and the donor oxygen constrained to its ground state value,\cite{Aquino2005} as depicted in the inset of Fig. \ref{fig:PT}a.
	
	Outside the cavity, photoexcitation induces the proton to transfer from the donor oxygen atom (\ch{O_D}) to the acceptor oxygen atom (\ch{O_A}). To characterize this proton transfer reaction, Fig. \ref{fig:PT}a shows the LR-TDDFT potential energy surface for the \ch{S0} and \ch{S1} states (black lines) when the proton is displaced by $\Delta R$ from its initial position along the vector connecting this initial position and \ch{O_A}. Intramolecular proton transfer is only facile on the \ch{S1} excited-state surface because for the \ch{S1} surface, the energy at $\Delta R =0$ is greater than the energies over the range $0< \Delta R < 0.45$ \AA.  Note that this figure depicts only a slice of the molecular multidimensional potential energy surface along $\Delta R$ and may not represent the reaction path followed by the proton.
	
	Next we consider the case when the \ch{S0$\rightarrow$S1} electronic transition at $\Delta R = 0$ is resonantly coupled to a cavity mode. In the singly excited manifold, at each proton displacement $\Delta R$, the bare \ch{S1} state is coupled to the \ch{S0} state dressed by an excited cavity mode (brown line). Quantitatively, we can express the effective light-matter Hamiltonian in the singly excited manifold as
	\begin{equation}\label{eq:H_polariton}
		\begin{pmatrix}
			E_{\text{S}_1}(\Delta R) & -\vmu_{\text{S}_0\rightarrow\text{S}_1}(\Delta R)\cdot\mathcal{E} \\
			-\vmu_{\text{S}_0\rightarrow\text{S}_1}(\Delta R)\cdot \mathcal{E} & E_{\text{S}_0}(\Delta R)+ E_{\text{ph}}
		\end{pmatrix}.
	\end{equation}
	Here, $E_{\text{S}_1}(\Delta R)$ and $E_{\text{S}_0}(\Delta R)$ denote the bare \ch{S1} and \ch{S0} state energies outside the cavity,  $\vmu_{\text{S}_0\rightarrow\text{S}_1}(\Delta R)$ denotes the transition dipole moment vector between these two states, and $\mathcal{E}$ denotes the effective electric field vector inside the cavity. Setting $\mathcal{E} = (0, 0.3 \text{\ a.u.}, 0)$, we diagonalize the Hamiltonian in Eq. \eqref{eq:H_polariton} at different values of $\Delta R$ and obtain the eigenstates, which correspond to the LP and UP states (red lines). Similar to a previous theoretical study \cite{Galego2016} of the ESC effect on photoisomerization, these calculations suggest that the proton transfer dynamics will be suppressed on both the LP and the UP potential energy surfaces compared to the original \ch{S1} channel.
	
	Analogous to the procedure applied to HCN above, we used the RT-NEO-TDDFT method to compute the power spectra when all electrons and the transferring proton are treated quantum mechanically. Fig. \ref{fig:PT}b shows the spectra  outside the cavity (blue line)  and inside a lossless cavity (red line) when the $y$-polarized cavity mode is resonantly coupled to the \ch{S0$\rightarrow$S1} electronic transition (blue peak at 3.295 eV) with $\varepsilon = 4 \times 10^{-3}$ a.u. Inside the cavity, a  pair of polaritons forms with the Rabi splitting $\Omega_\text{R} = 0.361$ eV (red line). \red{Note that when computing the spectra, a very weak pulse is used to excite the system, and therefore the molecule remains predominantly in the ground vibronic state and proton transfer does not occur.}
	
	To simulate photoinduced ESIPT in this molecule,  we model the \ch{S0$\rightarrow$S1} electronic transition  by exciting the electronic density matrix from the highest occupied molecular orbital (HOMO) to the lowest unoccupied molecular orbital (LUMO) at $t = 0$. \red{Inside the cavity, such an initial condition corresponds to an approximately equal excitation of the LP and UP, which could be realized experimentally by sending a wide-band pulse to excite the molecular system.} The proton transfer process is monitored by plotting the distance between the proton and \ch{O_D} and the distance between the proton and \ch{O_A}, where the proton moves away from the donor and toward the acceptor, as shown in \ref{fig:PT}c. These distances are computed with the expectation value of the proton position. The proton transfer time is defined as the time at which these two distances are the same. Outside the cavity (blue lines), the proton transfers at $\sim $ 4.2 fs and starts returning to \ch{O_D} at $t \approx 8$ fs, similar to previous RT-NEO-TDDFT studies of this molecule in free space.\cite{Zhao2020} 
	
	Inside the cavity (red lines), the proton dynamics is only  slightly suppressed. As the Rabi splitting here is similar to the splitting between the polaritonic potential energy surfaces ($\Delta R = 0$ in Fig. \ref{fig:PT}a), such a negligible cavity effect appears to conflict with  previous theoretical predictions. \cite{Galego2016} We hypothesized that this negligible effect was due to the faster timescale of proton transfer ($\sim $ 4 fs) compared to the Rabi oscillation timescale (11.5 fs based on $\Omega_\text{R} = 0.361$ eV). In this case, proton transfer itself effectively serves as a strong lossy channel that breaks down strong coupling. In other words, the energy transfer dynamics between the proton and the cavity mode, which is characterized by the Rabi splitting, is slower than the proton transfer dynamics, preventing the cavity from significantly influencing the proton transfer dynamics. According to this hypothesis, if the proton transfer dynamics is slowed down while the Rabi splitting remains the same, we expect to observe a stronger cavity suppression of proton transfer.
	
	To test this hypothesis, we increase the distance between the \ch{O_D} and \ch{O_A} atoms for the fixed geometry used in these simulations.  As shown in Figs. \ref{fig:PT}d,e, outside the cavity (blue lines), the proton transfer time increases to 7 fs and starts returning at 10 fs. When the lossless $y$-polarized cavity is again resonantly coupled to the \ch{S0$\rightarrow$S1} transition (blue peak at 3.611 eV) with $\varepsilon = 4 \times 10^{-3}$ a.u., the Rabi splitting ($\Omega_\text{R} = 0.348$ eV, \red{corresponding} to 11.9 fs) is similar to that in Fig. \ref{fig:PT}b, but the proton transfer dynamics (red lines) is greatly suppressed compared with that outside the cavity (blue lines). Comparing Fig. \ref{fig:PT}e to Fig. \ref{fig:PT}c, we observe a much stronger cavity suppression of proton transfer when the proton transfer timescale is extended while the Rabi splitting remains similar, thus confirming our above hypothesis. \red{For Fig. \ref{fig:PT}e, the corresponding electronic density (the green isosurface) and protonic density (the blue isosurface) at time $t = 0$ fs and 9.7 fs are plotted in the table of contents (TOC) figure.} When the  distance between the \ch{O_D} and \ch{O_A} atoms is further increased, as shown in Figs. \ref{fig:PT}f,g, proton transfer does not occur either inside or outside the cavity. 
	
	For the nuclear geometries used to generate the data in Fig. \ref{fig:PT}c or \ref{fig:PT}e, we have also observed that the cavity suppression effect becomes more substantial when the light-matter coupling strength $\varepsilon$ is increased (\red{see SI Fig. S2}). This finding is consistent with our above hypothesis because increasing $\varepsilon$ can lead to a smaller Rabi oscillation timescale due to a larger Rabi splitting. In this case, the proton transfer timescale becomes more similar to the Rabi oscillation timescale, which, according to our hypothesis, implies a larger cavity suppression effect for proton transfer. \red{Our hypothesis is also consistent with the observation that adding a cavity lifetime of 10 fs, which is slower than the proton transfer lifetime, does not alter the proton transfer dynamics (see SI Fig. S3).} 
	\begin{figure*}
		\centering
		\includegraphics[width=1.0\linewidth]{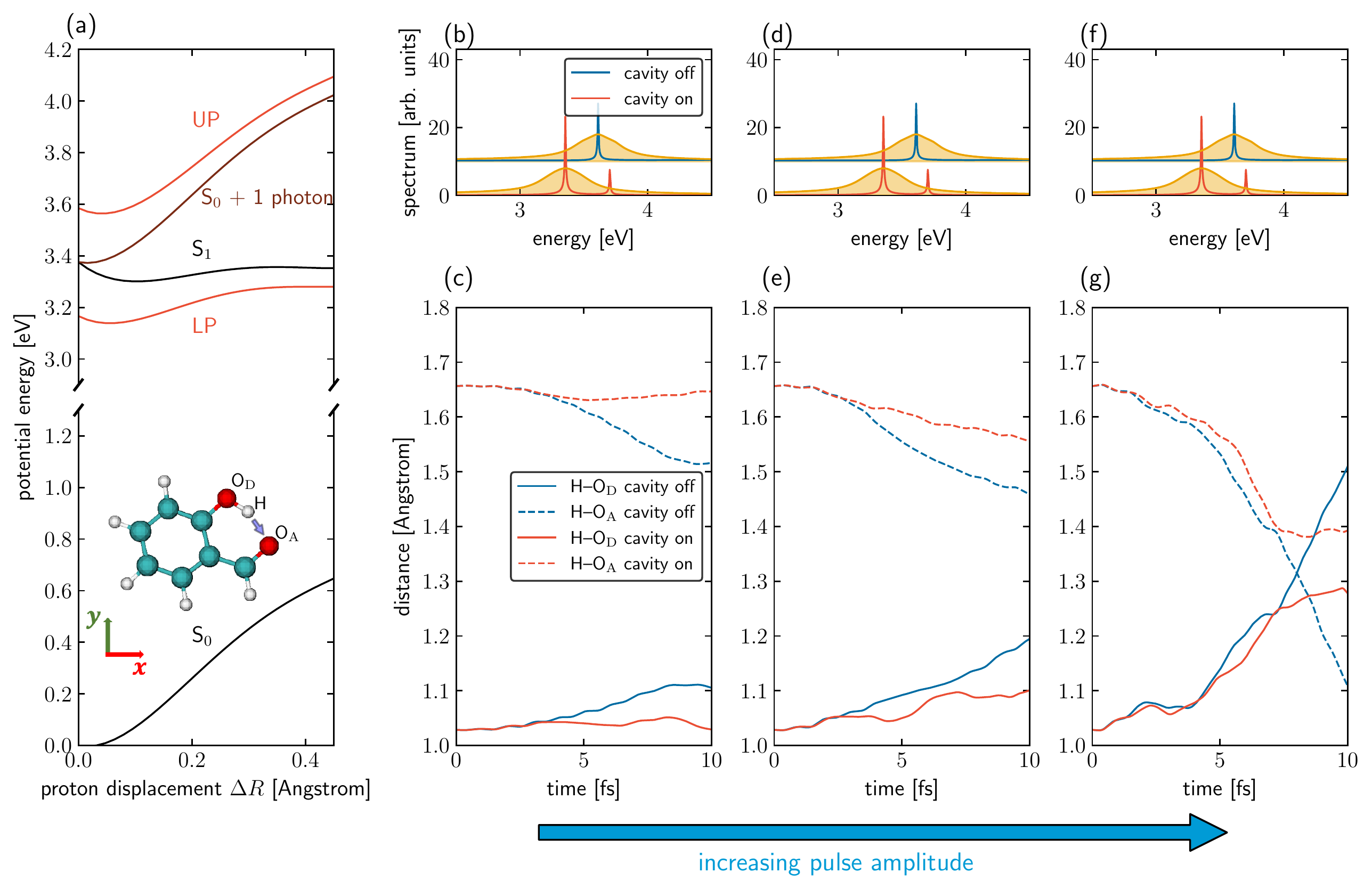}
		\caption{\red{Excited state intramolecular proton transfer dynamics for an oHBA molecule under an external pulse excitation. Part (a) is identical to Fig. \ref{fig:PT}a. For (b)-(g), the oHBA geometry is fixed to the same geometry as that used in Figs. \ref{fig:PT}(d,e), and therefore the spectra in (b), (d), and (f) are identical. In contrast to Fig. \ref{fig:PT}, where proton transfer is triggered by a HOMO to LUMO transition at time $t=0$, here proton transfer is induced by an external pulse excitation of the molecule (both inside and outside the cavity) with the form $\vE_{\text{ext}}(t) = E_0 \exp(-t^2/\sigma^2)\cos(\omega t) \ve_y$. In these simulations, $\sigma$ = 9.7 fs, $\omega$ = 3.611 eV outside the cavity (corresponding to the $S_0 \rightarrow S_1$ transition frequency), and $\omega$ = 3.355 eV inside the cavity (corresponding to the LP frequency). Moreover, $E_0 = 2\times 10^{-2}$ a.u. for  (c), $4\times 10^{-2}$ a.u. for  (e), and $8\times 10^{-2}$ a.u. for  (g). All other parameters are the same as in Fig. \ref{fig:PT}. The lineshapes of the external pulse  in the frequency domain are plotted as yellow shaded curves together with the spectra in the upper panel. Inside the cavity, the pulse mainly excites the LP, and the cavity suppresses the proton transfer dynamics.}}
		\label{fig:PT_pulse}
	\end{figure*}
	
	\red{A more realistic photoinduced ESIPT can be simulated by using an external Gaussian pulse to excite the molecular subsystem (not the cavity mode), in contrast to Fig. \ref{fig:PT}, where the proton transfer is triggered via a HOMO to LUMO transition at $t=0$. Analogous to Fig. \ref{fig:PT}, Fig. \ref{fig:PT_pulse} shows the proton transfer dynamics for oHBA under a Gaussian pulse excitation with the form $\vE_{\text{ext}}(t) = E_0 \exp(-t^2/\sigma^2)\cos(\omega t) \ve_y$. The pulse width is set as $\sigma = 9.7$ fs. Outside the cavity, the pulse frequency is chosen to peak at the $\text{S}_0\rightarrow \text{S}_1$ transition frequency ($\omega = 3.611$ eV), whereas inside the cavity, the pulse frequency is chosen to peak at the LP frequency ($\omega = 3.355$ eV).  The yellow shaded curves in the upper panel depict the corresponding pulse lineshapes in the frequency domain. For Figs. \ref{fig:PT_pulse}(b)-(g), the geometry of the oHBA molecule is fixed to the same geometry used in Figs. \ref{fig:PT}(d,e) and the amplitude of the pulse is increased from $E_0 = 2\times 10^{-2}$ a.u. (Fig. \ref{fig:PT_pulse}b,c) to $4\times 10^{-2}$ a.u. (Fig. \ref{fig:PT_pulse}d,e) to $8\times 10^{-2}$ a.u. (Fig. \ref{fig:PT_pulse}f,g). For all three pulse amplitudes, we observed a cavity suppression of proton transfer, in agreement with Fig. \ref{fig:PT}. 
	
	Our assumption that the pulse excites only the molecular subsystem (not the cavity mode) ensures similar initial conditions for the outside versus inside cavity situations, allowing a clean comparison between the inside versus outside cavity results. In reality, however, because plasmonic cavities usually have a much larger effective transition dipole moment than the confined molecule, the external pulse should interact more strongly with the cavity mode. Such a cavity enhancement of the external field amplitude should promote proton transfer, whereas the formation of polaritons, as we have observed in Fig. \ref{fig:PT}, is more likely to suppress proton transfer. These two effects clearly work against each other. Hence, in order to better simulate photoinduced proton transfer inside a plasmonic cavity, one should also take into account the cavity enhancement of the external field. Although  our code can be easily extended to include this enhancement (i.e., by assigning a very large effective transition dipole moment to the cavity mode), we do not report such a simulation here as it is beyond the scope of this work.}

	\section{5. Conclusion}
	
	In this manuscript, we have introduced the semiclassical RT-NEO-TDDFT approach for studying the real-time dynamics of molecular polaritons. By treating electrons and specified nuclei quantum mechanically with RT-NEO-TDDFT and propagating the coupled dynamics between the quantum molecular subsystem and the classical cavity mode(s) self-consistently, this approach not only  provides a unified description of ESC and VSC, but also can be used to probe the cavity effect on the coupled nuclear-electronic dynamics. Our \red{application} of this approach to excited state intramolecular proton transfer under single-molecule strong coupling \red{is generally consistent with the previous work \cite{Galego2016} showing that ESC can lead to suppression of photochemical reactions. However, our work reveals a new  consideration for the investigation of cavity effects} on chemical reactions: under single-molecule strong coupling for only the reactant, the cavity does not play a significant role when the chemical reaction timescale is faster than the Rabi oscillation timescale. In this case, the chemical reaction itself provides a strong lossy channel that may destroy strong coupling. As these timescales cannot be directly observed from the polaritonic potential energy surfaces, this work also highlights the importance of propagating real-time dynamics instead of considering only energetics. Furthermore, this  semiclassical approach provides the foundation for exploring collective strong coupling in chemical systems within optical cavities.

	%\section{Author Contributions}
	%T.E.L. and S.H.-S. designed research, T.E.L. performed research, T.E.L. and S.H.-S. analyzed data and wrote the paper.
	
	\section{Supporting information}
	
	Q-Chem input files and molecular geometries for the reported results; figure of the polariton spectrum for a single \ch{HCN} molecule under ESC when the light-matter coupling strength or the cavity loss rate is tuned\red{; figures of the proton transfer dynamics with a large Rabi splitting or with a cavity loss lifetime of 10 fs.}
	\red{The input files and plotting scripts are also available at Github (\url{https://github.com/TaoELi/semiclassical-rt-neo})}.
	
	\begin{acknowledgement}
	
    This material is based upon work supported by the Air Force Office of Scientific Research under AFOSR Award No. FA9550-18-1-0134. We thank John Tully,
    Alexander Soudakov, Jonathan Fetherolf, Qi Yu, Chris Malbon, and Mathew Chow for
    useful discussions.
    
    \end{acknowledgement}
	
	%\bibliography{references.bib}
	
	\providecommand{\latin}[1]{#1}
	\makeatletter
	\providecommand{\doi}
	{\begingroup\let\do\@makeother\dospecials
		\catcode`\{=1 \catcode`\}=2 \doi@aux}
	\providecommand{\doi@aux}[1]{\endgroup\texttt{#1}}
	\makeatother
	\providecommand*\mcitethebibliography{\thebibliography}
	\csname @ifundefined\endcsname{endmcitethebibliography}
	{\let\endmcitethebibliography\endthebibliography}{}

	\pagenumbering{gobble}
	
	

\section*{Supplementary material}

# Supporting Information

# Semiclassical Real-Time Nuclear-Electronic Orbital Dynamics for Molecular Polaritons: Unified Theory of Electronic and Vibrational Strong Couplings

Tao E. Li,* Zhen Tao, and Sharon Hammes-Schiffer*

*Department of Chemistry, Yale University, New Haven, Connecticut, 06520, USA*

E-mail: tao.li@yale.edu; sharon.hammes-schiffer@yale.edu

# 1. Q-Chem input file for HCN under VSC

──────────── HCN.in for Fig. 2c ────────────

```
$molecule
0 1
C  0.0 0.0 -0.5026771429
N  0.0 0.0  0.6555628571
H  0.0 0.0 -1.5728771429
$end

$rem
sym_ignore = 1
input_bohr = false
method = b3lyp
neo = true
neo_epc = epc172
basis = cc-pvdz
SCF_ALGORITHM = GDM
thresh = 14
s2thresh = 12
SCF_CONVERGENCE = 11
NEO_N_SCF_CONVERGENCE = 11
MAX_SCF_CYCLES = 500
NEO_PURECART = 1111
NEO_E_CONV = 12
MEM_TOTAL = 7000
NEO_VPP = 0
NEO_BASIS_LIN_DEP_THRESH = 8
$end

$neo_basis
H    3
```

```
S    1    1.000000
   2.828400 1.0
S    1    1.000000
   4.0 1.0
S    1    1.000000
   5.6569 1.0
S    1    1.000000
   8.0 1.0
S    1    1.000000
   11.3137 1.0
S    1    1.000000
   16.0 1.0
S    1    1.000000
   22.6274 1.0
S    1    1.000000
   32.0 1.0
P    1    1.000000
   2.828400 1.0
P    1    1.000000
   4.0 1.0
P    1    1.000000
   5.6569 1.0
P    1    1.000000
   8.0 1.0
P    1    1.000000
   11.3137 1.0
P    1    1.000000
   16.0 1.0
P    1    1.000000
   22.6274 1.0
P    1    1.000000
   32.0 1.0
D    1    1.000000
```

```
    2.828400 1.0
D    1      1.000000
    4.0 1.0
D    1      1.000000
    5.6569 1.0
D    1      1.000000
    8.0 1.0
D    1      1.000000
    11.3137 1.0
D    1      1.000000
    16.0 1.0
D    1      1.000000
    22.6274 1.0
D    1      1.000000
    32.0 1.0
$end


$neo_tdks
dt   0.04
maxiter 51677
field_type delta
field_amp 0.3
in_cavity true 0.3475 0 4e-4 1e8
rt_thresh 4
$end
```

---

The above input file can generate the data in Fig. 2c. The *$neo_tdks* section controls the RT-NEO-TDDFT dynamics. While all of the other input parameters are self-explanatory, "*in_cavity true 0.3475 0 4e-4 1e8*" indicates coupling the molecule to a single-mode cavity with a frequency of 0.3475 a.u., polarization direction of $x$ (0 to $x$, 1 to $y$, and 2 to $z$), coupling strength $\varepsilon = 4 \times 10^{-4}$ a.u., and lifetime $1/\gamma_c = 10^8$ a.u. Here, such a long lifetime

approximately means no cavity loss. "*rt_thresh 4*" indicates that the threshold of the predictor-corrector algorithm (the *eps* parameter in Algorithm 1 of Ref.[S1]) is $10^{-4}$.

## 2. Q-Chem input file for oHBA under ESC

oHBA.in for Fig. 4c

```
$molecule
0 1
C     -1.310008   1.258755   0.000000
C      0.019289   0.780580   0.000000
C      0.322586  -0.621636   0.000000
C     -0.761283  -1.465586   0.000000
C     -2.125342  -0.985468   0.000000
C     -2.398971   0.362640   0.000000
O      1.008708   1.658363   0.000000
C      1.725826  -1.062678   0.000000
O      2.682115  -0.219090   0.000000
H     -3.414260   0.733314   0.000000
H     -0.596742  -2.537818   0.000000
H     -2.926679  -1.713458   0.000000
H     -1.459388   2.331114   0.000000
H      1.924136  -2.138192   0.000000
H      1.844187   1.135044   0.000000
Gh     1.981048   0.913868   0.000000
Gh     2.117910   0.692694   0.000000
Gh     2.366459   0.659685   0.000000
$end


$rem
sym_ignore = 1
input_bohr = false
```

```
method = b3lyp
SCF_ALGORITHM = diis
MAX_SCF_CYCLES=200
basis = mixed
PURECART 1
SCF_CONVERGENCE = 8
mem_total = 7000
neo = true
NEO_E_CONV = 9
NEO_EPC = epc172
NEO_VPP = 0
$end

$basis
C 1
cc-pvdz
C 2
cc-pvdz
C 3
cc-pvdz
C 4
cc-pvdz
C 5
cc-pvdz
C 6
cc-pvdz
O 7
cc-pvdz
C 8
cc-pvdz
O 9
cc-pvdz
H 10
```

```
cc-pvdz

H 11

cc-pvdz

H 12

cc-pvdz

H 13

cc-pvdz

H 14

cc-pvdz

H 15

cc-pvdz

H 16

cc-pvdz

H 17

cc-pvdz

H 18

cc-pvdz

$end


$neo_basis

H     15

S    1    1.000000

4.0 1.0

P    1    1.000000

4.0 1.0

H     16

S    1    1.000000

4.0 1.0

P    1    1.000000

4.0 1.0

H     17

S    1    1.000000

4.0 1.0
```

```
P    1    1.000000
4.0 1.0
H    18
S    1    1.000000
4.0 1.0
P    1    1.000000
4.0 1.0
$end



$neo_tdks
electronic_HOMO2LUMO true
dt   0.04
maxiter 20000
field_type delta
field_amp 0e-2
in_cavity true 3.295 1 4e-3 1e8
rt_thresh 3
$end
```

The above input file can generate the data in Fig. 4c. In the *$neo_tdks* section, "*electronic_HOMO2LUMO true*" enforces a HOMO to LUMO transition in the electronic density matrix at $t = 0$ (but by default this control is turned off).

The geometry of the oHBA molecule in Fig. 4e is

oHBA.xyz for Fig. 4e

```
C  -1.301629  1.264863  0.000000
C   0.025677  0.802315  0.000000
C   0.308350 -0.595173  0.000000
C  -0.771161 -1.469584  0.000000
C  -2.105932 -1.005452  0.000000
```

```
C -2.364799  0.358631 0.000000

O  1.014139  1.694260 0.000000

C  1.693213 -1.060458 0.000000

O  2.668602 -0.278478 0.000000

H -3.383290  0.724581 0.000000

H -0.580558 -2.537265 0.000000

H -2.918862 -1.719777 0.000000

H -1.469874  2.334074 0.000000

H  1.853545 -2.147310 0.000000

H  1.860084  1.188551 0.000000
```

---

The additional three protonic basis function centers are the same as those in the above oHBA Q-Chem input file.

The geometry of the oHBA molecule in Fig. 4g is

---- oHBA.xyz for Fig. 4g ----

```
C -1.293251  1.270970 0.000000

C  0.032066  0.824049 0.000000

C  0.294115 -0.568710 0.000000

C -0.781038 -1.473582 0.000000

C -2.086523 -1.025436 0.000000

C -2.330627  0.354621 0.000000

O  1.019570  1.730158 0.000000

C  1.660600 -1.058238 0.000000

O  2.655089 -0.337865 0.000000

H -3.352319  0.715847 0.000000

H -0.564373 -2.536711 0.000000

H -2.911046 -1.726097 0.000000

H -1.480360  2.337034 0.000000

H  1.782954 -2.156428 0.000000

H  1.875980  1.242058 0.000000
```

The additional three protonic basis function centers are the same as those in the above oHBA Q-Chem input file.

For Fig. 5, in order to simulate proton transfer under an external Gaussian pulse excitation, we need to slighly change the $neo\_tdks$ input parameters:

──────────────── `$neo_tdks$` input parameters for Fig. 5 ────────────────

```
$neo_tdks
electronic_HOMO2LUMO false
dt   0.04
maxiter 20000
field_type gaussian 0.0 400.0 3.355
field_amp 8e-2
field_direction 1
in_cavity true 3.611 1 4e-3 1e8
in_cavity_excite_molecule true
rt_thresh 3
$end
```

Here, "*field_type gaussian 0.0 400.0 3.355*", "*field_amp 8e-2*", and "*field_direction 1*" collectively define an external Gaussian pulse with the following form: $\mathbf{E}_{\text{ext}}(t) = E_0 \exp[-(t-t_0)^2/\sigma^2]\cos(\omega t)\mathbf{e}_y$, where the field amplitude is $E_0 = 8 \times 10^{-2}$ a.u., $t_0 = 0$ a.u., $\sigma = 400.0$ a.u., and frequency $\omega = 3.355$ eV. Without the parameter "*field_direction 1*", the field will interact with the molecule in three dimensions. "*in_cavity_excite_molecule true*" ensures that the external pulse interacts only with the molecule (not the cavity mode). By default, the external field interacts with only the cavity mode.

# 3. Additional simulation data

## 3.1. Polaritonic spectra under ESC

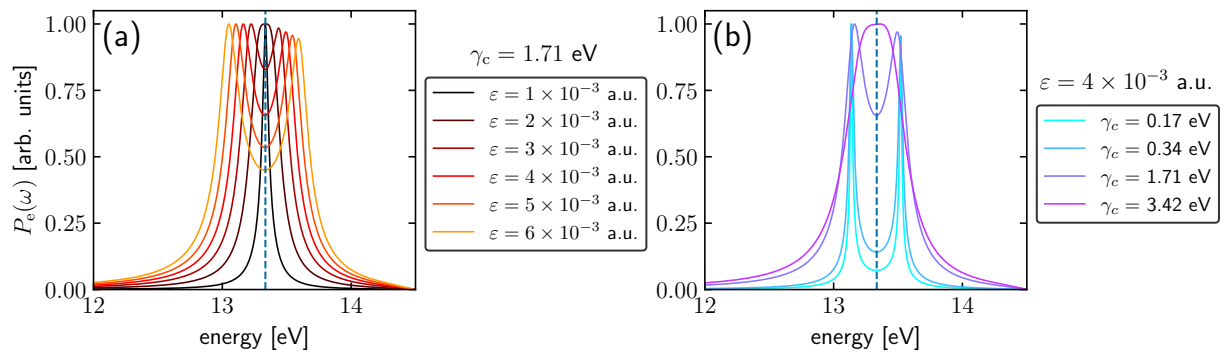

Figure S1: Power spectrum, $P_\text{e}(\omega)$, of the HCN electronic dipole moment when the HCN molecule is resonantly coupled to the $z$-polarized cavity with frequency $\omega_\text{c} = 13.334$ eV (vertical dashed blue lines). (a) The cavity loss is $\gamma_\text{c} = 1.71$ eV and the light-matter coupling strength ranges from $\varepsilon = 1 \times 10^{-3}$ a.u. (black line) to $6 \times 10^{-3}$ a.u. (orange line). Note that a gradual increase of Rabi splitting is observed when the coupling strength is amplified. (b) The cavity loss ranges from $\gamma_\text{c} = 0.17$ eV (cyan line) to 3.42 eV (purple line) with the light-matter coupling strength set to $\varepsilon = 4 \times 10^{-3}$ a.u. Increasing the cavity loss transforms the system from strong coupling (with a peak splitting) to weak coupling (with no peak splitting). All other simulation details are the same as for Fig. 1d. Since the intrinsic molecular linewidth is $1.7 \times 10^{-3}$ eV due to the small damping term $e^{-\gamma t}$ used for processing the real-time signals, all of the linewidths in this figure arise mainly from the coupling to the lossy cavity.

Fig. S1a demonstrates the electronic power spectrum inside a lossy cavity when the coupling strength is tuned from $1 \times 10^{-3}$ a.u. (black line) to $6 \times 10^{-3}$ a.u. (orange line) and all of the other simulation details are the same as in Fig. 1d. When the coupling strength is small ($\varepsilon \leq 2 \times 10^{-3}$ a.u.), since the cavity loss is large, the molecule is weakly coupled to the cavity mode and no peak splitting is observed. When the coupling strength increases, the Rabi splitting overcomes the cavity loss, and two polaritons are observed. Fig. S1b shows the effect of cavity loss on the molecular lineshape when the coupling strength is fixed as $\varepsilon = 4 \times 10^{-3}$ a.u. When the cavity loss increases from $\gamma_\text{c} = 0.17$ eV to 3.42 eV (cyan to purple), the peak splitting disappears.

## 3.2. Proton transfer under large light-matter coupling

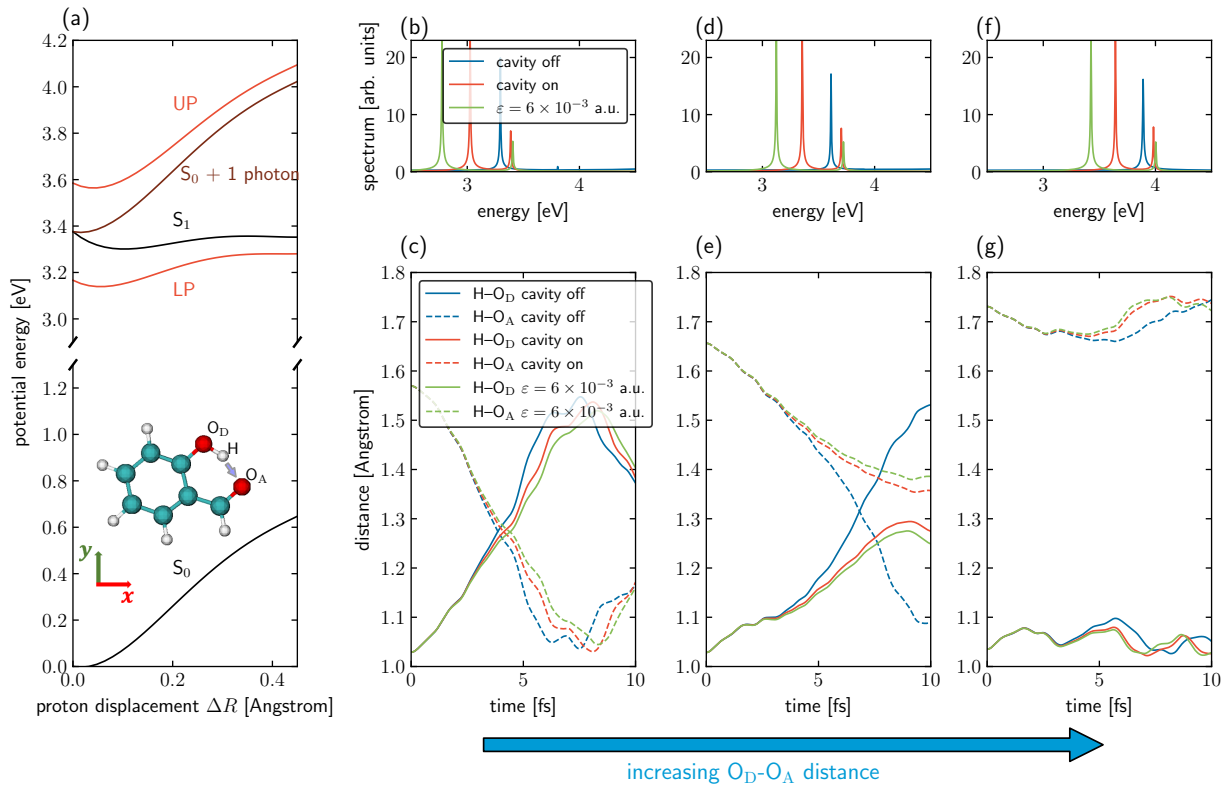

Figure S2: Another version of Fig. 4 in the main text. The only difference is that, from (b) to (g), the green lines represent the results corresponding to a large light-matter coupling $\varepsilon = 6 \times 10^{-3}$ a.u., where all other parameters are the same as the inside cavity results in Fig. 4 obtained with $\varepsilon = 4 \times 10^{-3}$ a.u. (red lines). Note that increasing the Rabi splitting can further suppress the proton transfer dynamics.

## 3.3. Proton transfer dynamics under a large cavity loss

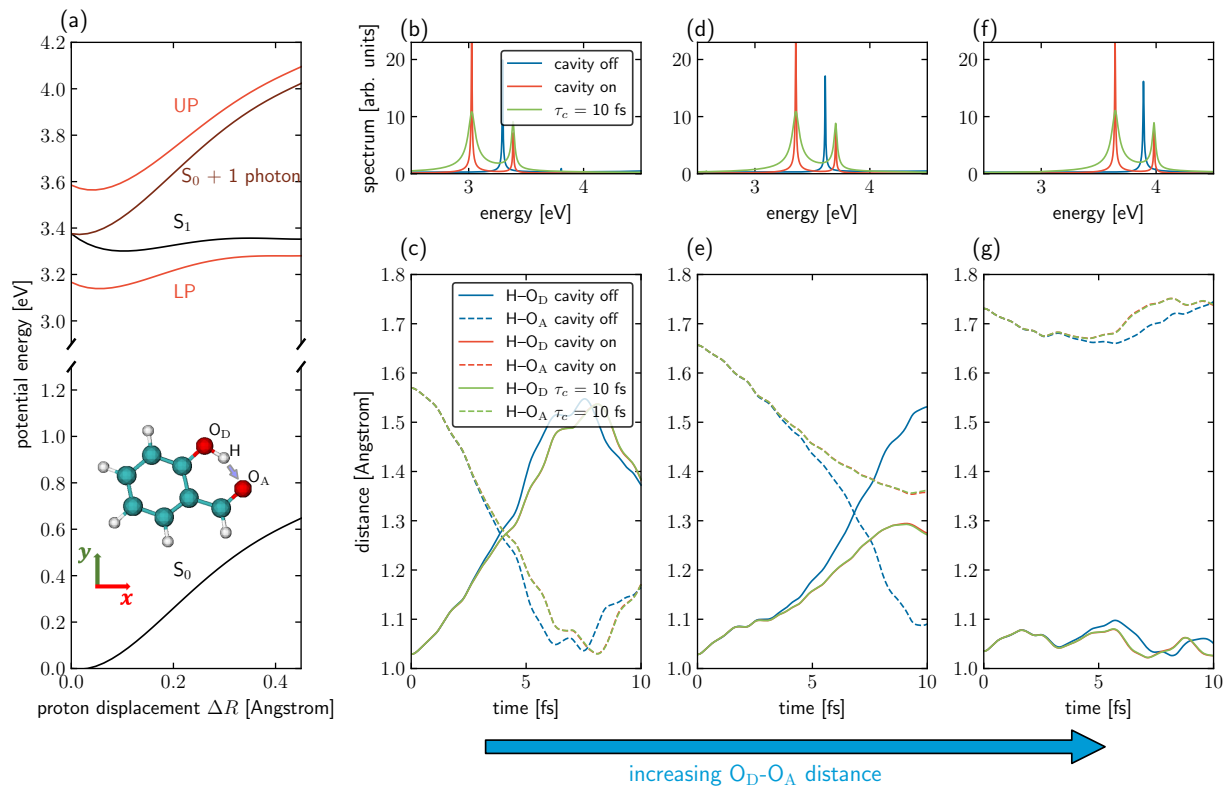

Figure S3: Another version of Fig. 4 in the main text. The only difference is that, from (b) to (g), the green lines represent the results corresponding to a large cavity loss $\tau_c = 1/\gamma_c = 10$ fs, where all other parameters are the same as the inside cavity results in Fig. 4 obtained with a lossless cavity (red lines). Note that adding a cavity loss broadens the polariton linewidths, but the proton transfer dynamics is not altered.

# Graphical TOC Entry

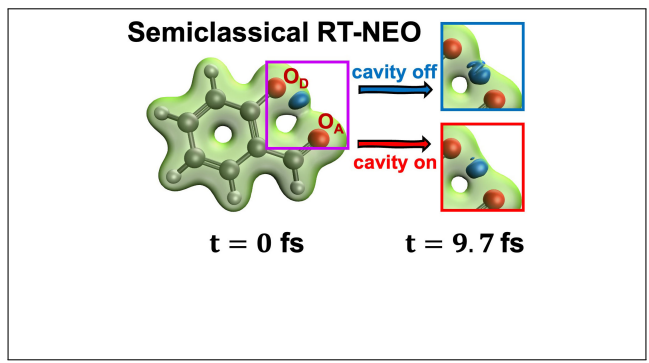


	
\end{document}